# Wigner function and photon number distribution of a superradiant state in semiconductor laser structures

Peter P. Vasil'ev[1,2*], Richard V. Penty[1], Ian H. White[3]

[1]Centre for Photonic Systems, University of Cambridge, 9 JJ Thomson Avenue, Cambridge, CB3 0FA, United Kingdom
[2]Quantum Radiophysics Department, PN Lebedev Physical Institute, 53 Leninsky Prospect, Moscow 119991, Russia
[3]University of Bath, Claverton Down, Bath, BA2 7AY, United Kingdom
[*]pv261@cam.ac.uk

For the visualization of quantum states, the approach based on Wigner functions can be very effective. Homodyne detection has been extensively used to obtain the density matrix, Wigner functions and tomographic reconstructions of optical fields for many thermal, coherent or squeezed states. Here, we use time-domain optical homodyne tomography for the quantum state recognition and reconstruction of the femtosecond optical field from a nonequilibrium superradiant coherent electron-hole state formed in a semiconductor laser structure. We observe severe deviations from the Poissonian statistics of the photons associated with the coherent laser state when the transformation from lasing to superradiance occurs. The reconstructed Wigner functions show large areas of negative values, a characteristic sign of non-classicality, demonstrating the quantum nature of the generated superradiant emission. The photon number distribution and Wigner function of the SR state are very similar to those of the displaced Fock state.

Recent advances in the field of quantum optics have paved a way to light-based quantum technologies, including quantum communications, quantum sensing, quantum simulation, cryptography and computation [1,2]. Quantum optics has enabled these new technologies, which promise radically new capabilities derived from design principles based on quantum mechanics rather than principles based on classical physics. Light plays a central role in all of these applications because it is an ideal medium for transmitting quantum information and is at the heart of many of the most promising and potentially transformative quantum technologies. The approaches and experimental techniques developed in quantum optics allow for investigating a number of unique phenomena such as entanglement and teleportation and studying different quantum states of field and matter, including coherent, squeezed, and Schrödinger-cat states [3].

For the visualization of quantum states, the approach based on Wigner functions has been widely used for a long time [4-6]. This representation reveals striking properties of different quantum states, for example, the oscillatory photon statistics of highly squeezed states or the possibility of reconstructing quantum states using optical tomography [7]. The Wigner function is essentially a phase space probability distribution function of the simultaneous values of x for the coordinates and p for the momenta [8]. As Wigner pointed out, this function is real but is not always positive definite. Wigner in his study was primarily interested in the corrections to the description of thermodynamic equilibrium. Among the first adopters of the Wigner function was optics when the quantum phase space distribution was used to describe the coherence and polarization of optical fields [9]. In quantum optics, the density matrix and the Wigner function of a quantum state can be reconstructed using optical coherent tomography which is often done with homodyne detection.

At the same time, the Dicke model [10] has recently drawn renewed interest because it is a relatively simple system in which one can find entanglement and related phenomena, and because it can be realized in a wider range of systems than in the original cavity quantum electrodynamics case. A new aspect emerged when it was realized that the superradiant (SR) quantum-phase transition is relevant to quantum information and quantum computing [11-13]. SR emission has been generated from many types of media, including quantum dots at cryogenic temperatures [14]. We have previously reported the observation of the SR phase transition in 2D and 3D semiconductor laser devices with different material compositions at room temperature [15-18]. The observed SR emission exhibited a number of remarkable features, including stronger coherence than lasing, superluminal propagation and ultrabright internal second-harmonic generation. The physical reason behind these effects consists in the quantum phase transition and non-equilibrium condensation of electrons and holes in phase space which occurs in the semiconductor [19]. Instead of an ensemble of individual e-h pairs, one has a coherent BCS-like collective state which exhibits a long-range order with two areas of macroscopic polarization [15,16,19]. Quantum properties of e-h ensembles transfer to the emitted SR light making it quantum. In this paper, we determine the density matrix, the photon number distribution and reconstruct the Wigner function of SR light emitted due to radiative recombination of the coherent e-h BCS-like collective state using optical quantum-state tomography exploiting homodyne detection.

Experimental set-up

A wide variety of multiple-section GaAs/AlGaAs laser structures capable of generating SR emission have been studied. In general, they have two gain and one saturable absorber sections of different geometries with different gain/absorber ratios and total cavity lengths between 90 and 250 microns. The devices under test are described in detail in our previous publications [16-18]. The varied composition of the GaAs/AlGaAs heterostructures results in a broad range of operating wavelengths in the range of 810 to 890 nm. All devices can operate under continuous-wave operation or SR regimes, depending on the driving conditions. Figure 1 shows the experimental set-up. The emission from a sample is collimated by a diode lens with an NA of 0.5 and is fed into an interferometer consisting of two non-polarizing beam splitters. The input beam is attenuated by neutral density filters with different optical densities depending on the required signal level on the detector. One mirror is attached to a piezoelectric translator and placed on a precise translation stage. The interferometer is aligned to provide a zero difference of the optical paths of both arms. The two beams from the interferometer are focused on the two inputs of a balanced photodetector Thorlabs PDB435C. The homodyne photocurrent is recorded by a digital oscilloscope with a bandwidth of 300 MHz and a sampling rate of up to 4 GSamples/s. The digitized output is numerically integrated over time intervals corresponding to the duration of the pulse. Each integral is associated with a single quadrature measurement. This method is used as a standard in pulsed homodyne detection using mode-locked solid-state and pulsed semiconductor laser sources [21,22].

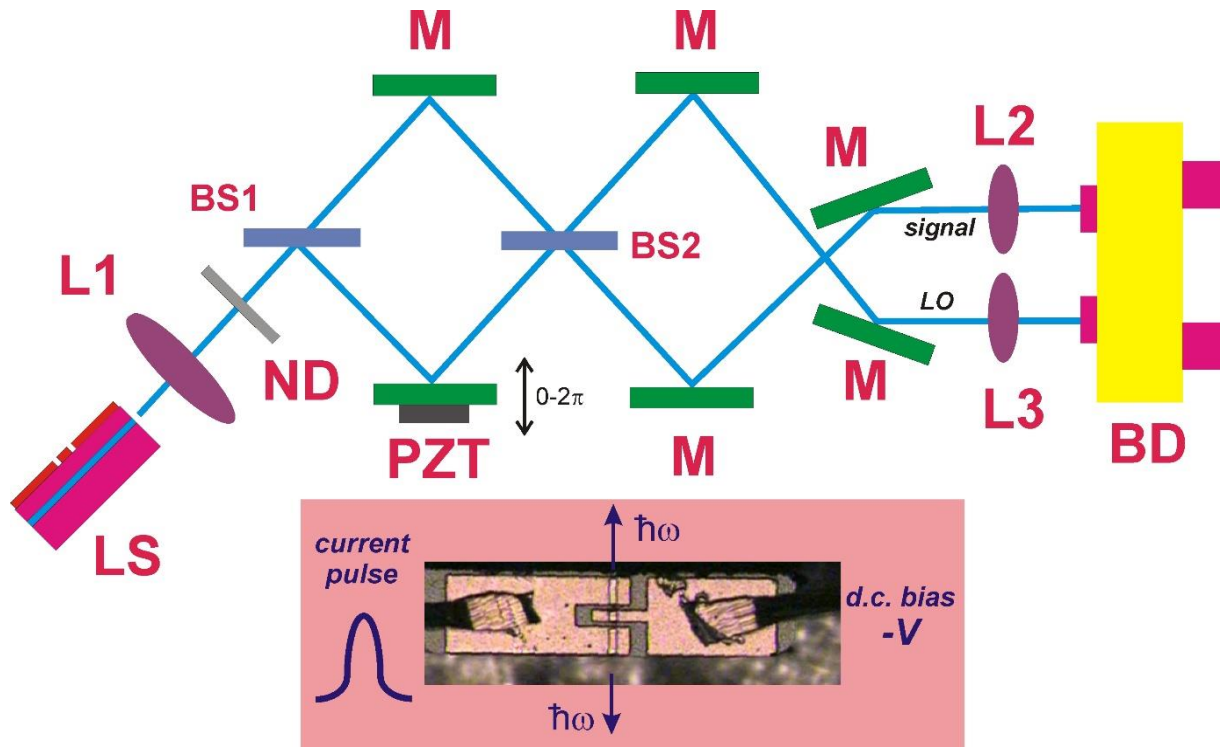

Figure 1. Experimental setup. LS is a laser sample, L1, L2, and L3 – lenses, ND – neutral density filters, BS1 and BS2 – non-polarizing beam splitters, M – mirrors, PZT - piezoelectric translator, BD – balanced photodetector. Inset: the microscopic top view of LS with a 3-section design. The nanosecond current pulses are applied on the left electrode, the right section is reverse biased by a d.c. voltage. The output light is emitted in opposite directions.

The present set-up is first tested for quantum state reconstruction using two single-mode c.w. and pulsed semiconductor lasers operating at 652 and 778 nm. The density matrix and Wigner functions are reconstructed for cases with different numbers of photons per mode. The test results provided standard coherent laser states with Poissonian statistics of photons as expected.

Results

When current pulses are applied on both laser contact simultaneously (see Fig. 1), the standard laser emission is generated. The lasing threshold is typically below 100 mA. The optical spectrum and the output light pulse are shown in Fig. 2 (left). The spectrum is almost single-mode one with a number of low-intensity longitudinal side modes. The optical pulse exhibits some typical relaxation oscillations. A strong reverse bias of the central absorber section (the electrode on the right in Fig. 1) allows for frustrating the onset of lasing in the sample for a few additional nanoseconds and achieving much greater e-h densities than the threshold lasing density. At a large enough e-h density SR phase transition occurs in the system and giant superradiant 200-400 femtosecond long pulses with peak powers of over 120-200 W are emitted [19, 20].

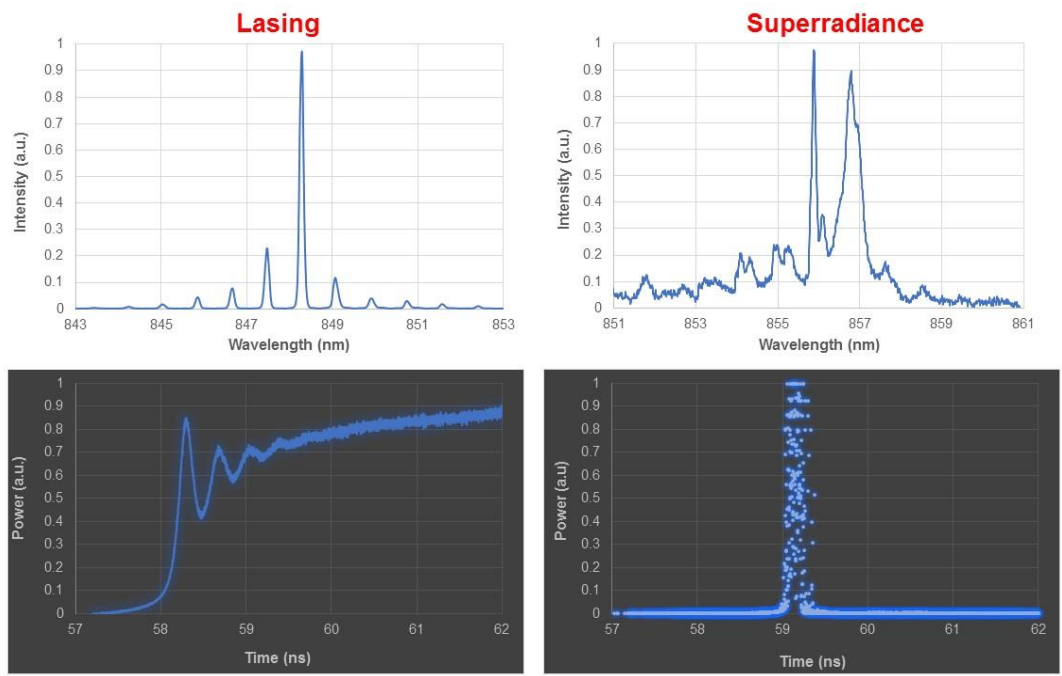

Figure 2. Optical spectra and output waveform as detected by a high-speed photodiode and sampling oscilloscope of lasing and SR emission from the same device.

The SR optical spectrum consists of a doublet, the emission peak being red-shifted by about 8 nm (Fig. 2). The separation of the components of the doublet exceeds by over 10% the spacing of the laser modes, which corresponds to a decrease of the group refractive index during SR [17]. The femtosecond SR pulse exhibits a huge jitter which is common to all SR pulses [23]. Figure 3 presents the results of the reconstruction of the coherent laser state at a certain current amplitude above the lasing threshold, where the nanosecond current pulses are applied to both device sections. This causes the uniform current injection without any saturable absorption. The homodyne traces are measured for the relative delay range corresponding to the central interference optical cycle. The optical spectrum and output light pulse are shown in Fig. 2.

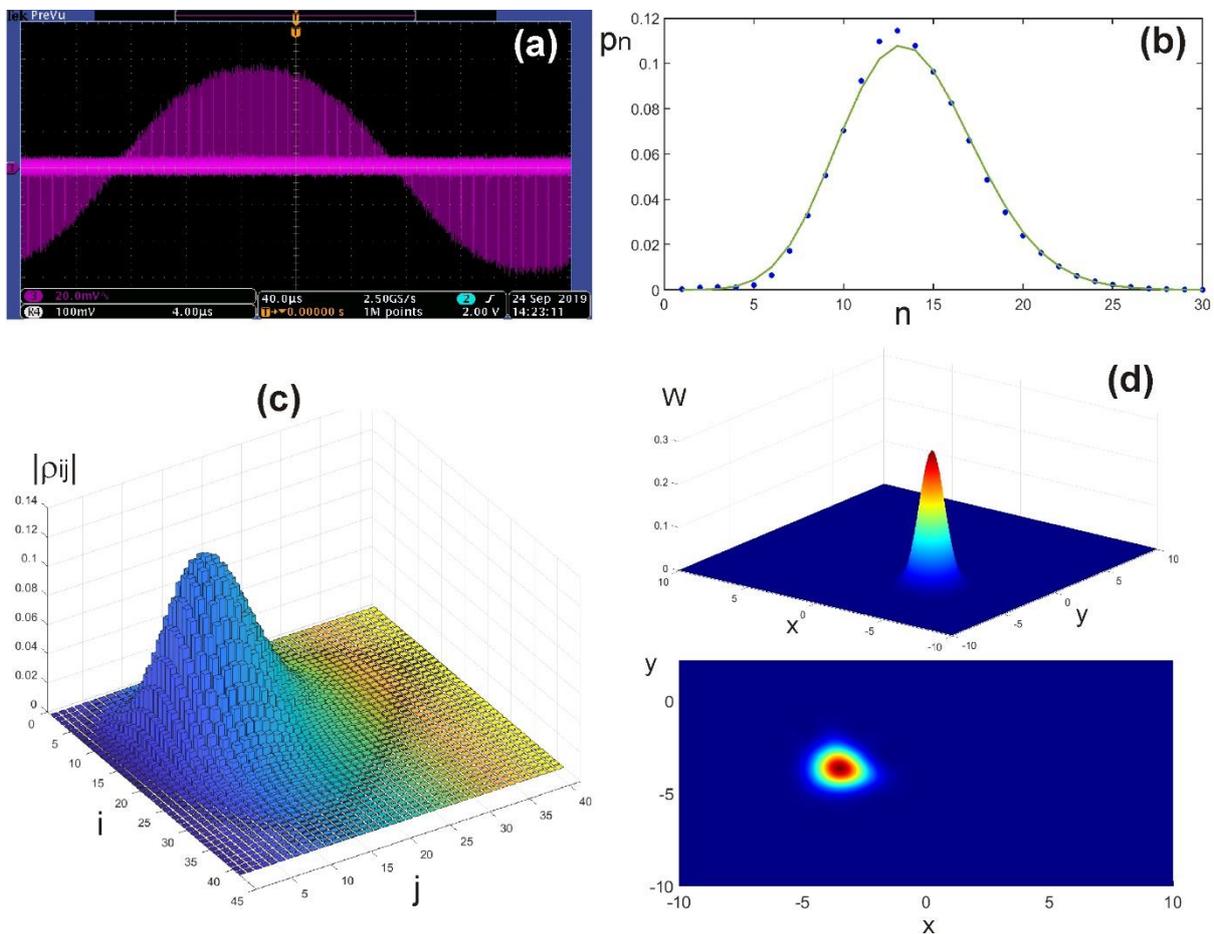

Figure 3. The reconstruction of the coherent laser state. (a) the oscilloscope quadrature, (b) photon number distribution $p_n$. The dots correspond to the measured data. The green line presents the Poissonian statistics of the photons, (c) 3D plot of the absolute values of the density matrix elements $|\rho_{ij}|$, (d) the reconstructed Wigner function and its 2D projection.

The green line in Fig. 3 (b) is the Poissonian statistics of the photons which is given by

$$p_n = \frac{\langle n \rangle^n}{n!} e^{-\langle n \rangle} \tag{1}$$

with $\langle n \rangle$ being the mean photon number. For the given experimental situation, we obviously have a coherent laser state obeying the Poissonian statistics with $\langle n \rangle \sim 14$. The experimental data allows for estimating the second-order correlation function $g^{(2)}(0)$

$$g^{(2)}(0) = \frac{\langle n^2 \rangle - \langle n \rangle}{\langle n \rangle^2} \tag{2}$$

The experimental value of $g^{(2)}(0)$ in Fig. 3 is 0.998, which is very close to the theoretical value of 1.0 for a coherent state.

Figures 4 and 5 present the results of the reconstruction of the SR states at different current amplitudes *I* and reverse biases *V* on the saturable absorber. The output pulses in time domain for all 3 SR states were similar to that presented in Fig. 2.

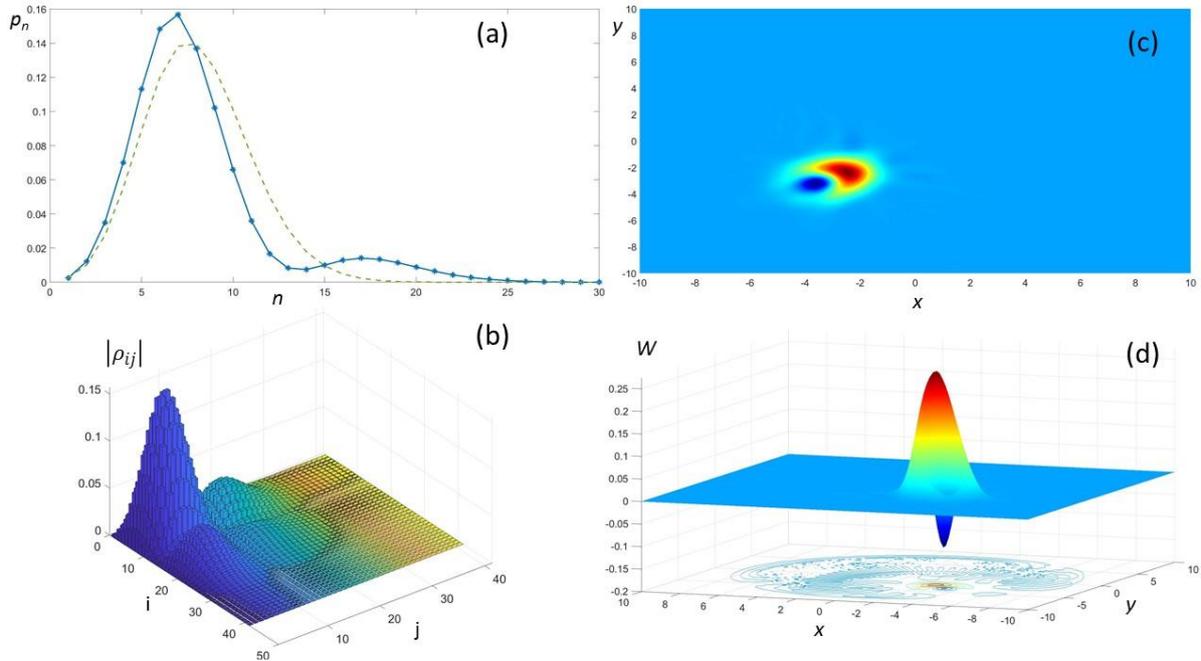

Figure 4. The reconstruction of the SR state at I = 410 mA and V = - 4.1 V. (a) photon number distribution $p_n$. The dots and blue line correspond to the measured data. The green dashed line presents the Poissonian statistics of photons of a coherent state, (b) 3D plot of the absolute values of the density matrix elements $|\rho_{ij}|$, (c) 2D projection of the reconstructed 3D Wigner function (d).

In contrast to the coherent laser state (Fig. 3), all SR states exhibit severe deviations of the photon statistics from the Poissonian statistics and significant areas where the Wigner functions have negative values. We see that the Wigner function of the SR states has negative values. This surprising feature makes it impossible to interpret the Wigner function as a true probability distribution. The photon number distribution in Fig. 4(a) and 5(a) has 2 peaks which are particularly visible in the last case.

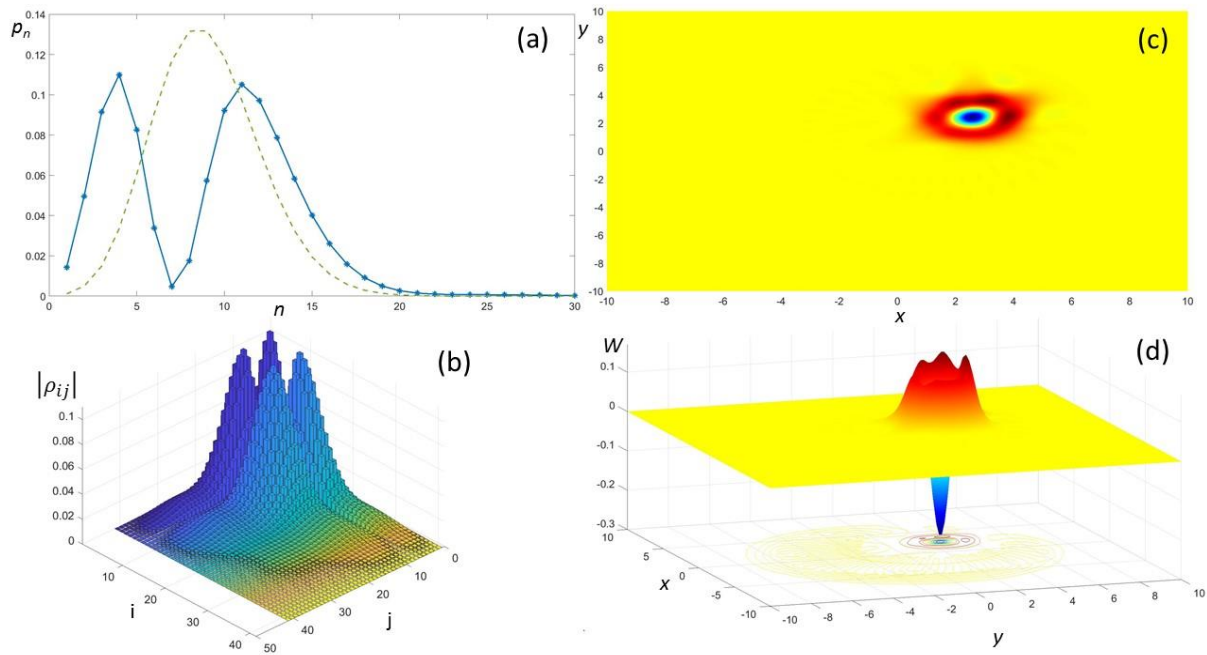

Figure 5. The reconstruction of the SR state at I = 630 mA and V = - 6.7 V. (a) photon number distribution $p_n$. The dots and blue line correspond to the measured data. The green dashed line presents the Poissonian statistics of photons of a coherent state, (b) 3D plot of the absolute values of the density matrix elements $|\rho_{ij}|$, (c) 2D projection of the reconstructed 3D Wigner function (d).

Similar results were obtained with a number of the SR devices. Further increase of the driving current amplitude above around 800 mA or the value of the reverse bias below – 9 V resulted in a catastrophic degradation of the samples. That happened due to either an electrical discharge in the gaps between the upper metal electrodes or an optical breakdown of the facets of the samples by the huge power flux density of the femtosecond SR pulses [24].

Discussion

The geometry of all samples with two gain sections at both ends and the saturable absorber at the centre (see Fig. 1) implies the presence of two areas of high e-h densities. Two areas of macroscopic polarization are formed there during the SR phase transition. They are separated by a narrow (~ 10 μm) saturable absorber section at the centre. As we reported previously, the coherent beating of the electric field with THz frequencies [15] and long-range order [16] have been observed. In contrast to lasing, when the optical field is coherent and the active medium is incoherent, the feature of SR is that a non-equilibrium coherent BCS-like state is formed, e-h pairs being collectively paired like in an ensemble of Cooper pairs [15]. One can expect that quantum coherence of this state is translated to the electromagnetic field of the emitted light. The oscillatory behaviour of the photon number distributions and the specific shapes of the Wigner functions in Fig. 4 and 5 suggest the observation of quantum interference between coherent states [25]. Quantum interference in phase space is associated with off-diagonal matrix elements of quantum states. The quantum interference results, in particular, in oscillation behaviour of the photon number distribution. In addition in our case, the interference between two components of the coherent e-h state, located at opposite ends of the sample, is a sensitive function of the relative phase difference and can generate different realizations of coherent electromagnetic field beatings in the time domain.

One can see that both photon number distribution and the Wigner function of the SR state shown in Fig. 5 have specific shapes which bear a resemblance to those of a displaced number state, which is often called as displaced Fock state (DFS) [25,26]. Displaced Fock states are non-classical generalizations of coherent states. The DFS provides a clear illustration of the concept of interference in phase space. The photon number distribution of a displaced Fock state |α, k> is given [25]

$$p_n^{(\alpha,k)} = \frac{e^{-|\alpha|^2}|\alpha|^{2(n-k)}}{n!\,k!} \left|\sum_{m=0}^{k} \frac{n!k!(-1)^m |\alpha|^{2(k-m)}}{m!(k-m)!(n-m)!}\right|^2 \qquad (3).$$

The Wigner function of the DFS is equal to the shifted Wigner function of the Fock state |k> [25]

$$W_k^{(\alpha)}(\beta) = \frac{2(-1)^k}{\pi} e^{(-2|\beta-\alpha|)^2} L_k(4|\beta-\alpha|^2) \qquad (4)$$

where $L_k(x)$ is the Laguerre polynomial of order $k$. Let us consider a special case of $k = 1$. According to Eq. (3), the function $p_n^{(\alpha,1)}$ of the DFS is equal to 0 for $n = |\alpha|^2$. This is opposite to a coherent state for which the photon number distribution reaches the maximum value at $n = |\alpha|^2$.

Compare now the reconstructed SR state with the displaced Fock state |α, k=1>. Figure 6 presents the characteristics of the SR state generated in one of the samples at I = 770 mA and V = - 7.2 V and the photon number distribution and the Wigner function of the DFS calculated using Eqs. (3)-(4). We use the notation $x = Re\ \beta$ and $y = Im\ \beta$. The photon number distribution of the SR state has a minimum value at n = 9. The calculated photon number distribution of the DFS fits very well to the experimental data for $|\alpha|^2 = 9$ (see Fig. 6(a)). The calculated Wigner function of the DFS looks a lot like the experimental one of the SR state. The latter is a bit broader than the former. The value of the negative peak of the Wigner function of the SR state is also smaller than that of the DFS. This may be caused by the fact that the calculations according to Eq. (4) do not take into account the damping and dissipation which are always present in real systems. Another reason may be determined by the ultrashort lifetime of the SR state under study.

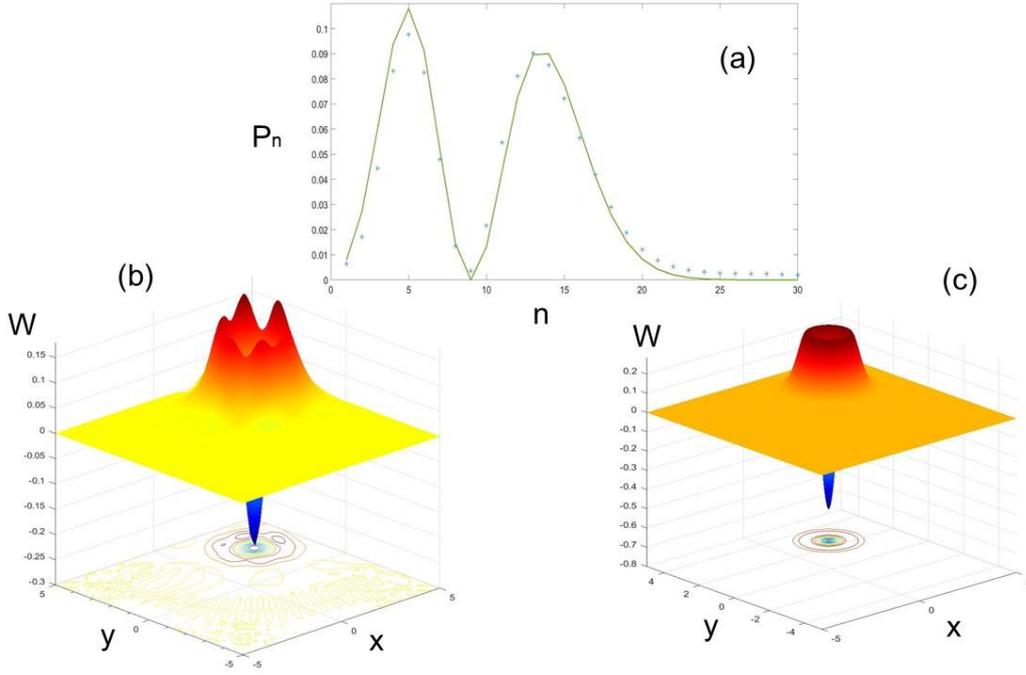

Figure 6. Comparison of the parameters of the measured SR state generated at I = 770 mA and V = - 7.2 V and the calculated displaced Fock state |α, k=1⟩. (a) The photon number distribution. The blue stars represent the SR state, the green line corresponds to the DFS with $|α|^2 = 9$; (b) The reconstructed Wigner function of the SR state; (c) the calculated Wigner function of the DFS with α = 2.15+2.1*$i$.

In conclusion, we experimentally demonstrate that quantum light can be generated from semiconductor laser structures during the superradiant phase transition. Using a time-domain optical homodyne tomography we have performed quantum state recognition and the reconstruction of the femtosecond optical field from a nonequilibrium superradiant coherent electron-hole state. We observe the oscillatory behaviour and severe deviations from the Poissonian statistics of the photons. We interpret the photon distribution oscillations as the result of quantum interference in phase space. The reconstructed Wigner functions of SR states show large areas of negative values which is a characteristic sign of non-classicality. The photon number distribution and Wigner function of the SR state are found to be very similar to those of the displaced Fock state |α, k⟩ with *k = 1*. One of the features of DFS is that the state is determined by its photon number while the phase is completely random [26]. This fact can explain abnormally large phase fluctuations of the SR emission which have been previously observed [23].

Supplementary Information

Further details on homodyne detection, quantum states recognition, and tests of coherent states using c.w. and pulsed semiconductor lasers.

# Supplementary Information

Wigner function and photon number distribution of a superradiant state in semiconductor laser structures

Peter P. Vasil'ev[1,2*], Richard V. Penty[1], Ian H. White[1]

[1]Centre for Photonic Systems, University of Cambridge, 9 JJ Thomson Avenue, Cambridge, CB3 0FA, United Kingdom

[2]Quantum Radiophysics Department, PN Lebedev Physical Institute, 53 Leninsky Prospect, Moscow 119991, Russia

## I. Experimental quadratures

The measured quantity is the output voltage from the balanced detector. Its output depends on the phase difference between the two arms of the interferometer. The phase difference $\phi$ is varied in the range [0, 2π] using the piezoelectric translator. The voltage proportional to the detector photocurrent is acquired by the digital oscilloscope. Each difference pulse is numerically integrated and associated with a single quadrature measurement. The collected experimental homodyne data $\{V_i\}$ is normalized for further processing by the constant δ

$$Y_i = \frac{V_i}{\delta}$$

with

$$\delta = \sqrt{2\langle V^2 \rangle_0}$$

where $\langle V^2 \rangle_0$ is the experimental voltage variance for the case when the signal beam is blocked [1].

## II. Density matrix and Wigner function reconstruction

We used the technique for the quantum state reconstruction based on the maximum-likelihood algorithm built with the use of the method described by Jeff Lundeen in http://www.photonicquantum.info/Tools.html. The Matlab code was written for the experimental data processing and the quantum state reconstruction. The input parameters were the phases $\{\phi_i\}$ and the corresponding normalized homodyne data $\{Y_i\}$. The Matlab program allowed for calculating the density matrix, the photon number distribution, and the Wigner function, and their comparison with those of coherent and displaced number states.

## III. Testing the experimental setup using c.w. and pulsed lasers

We present here the results of the testing of our homodyne detection setup. We reconstruct coherent states of two semiconductor lasers operating in the c.w. mode at 652 nm and in pulsed mode at 782 nm. Figures 1-3 presents the experimental quadrature, reconstructed density matrix and Wigner function of the c.w. laser emission at 652 nm and around 15 % excess of the driving current above the laser threshold.

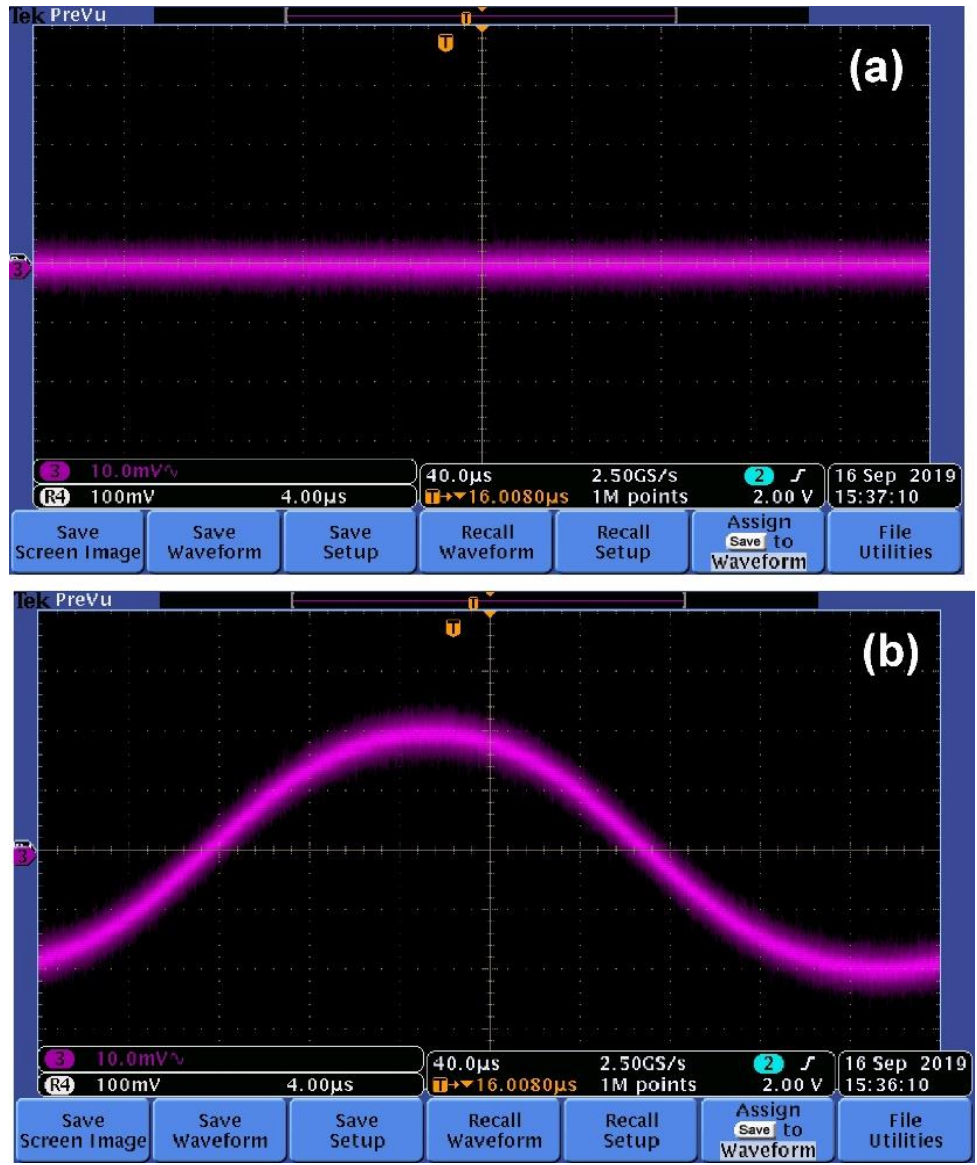

Figure 1. (a) the signal is blocked and (b) the homodyne quadrature.

Figure 4 presents the output waveform, the optical spectrum of the emission from the pulsed laser operating at 782 nm, and the corresponding homodyne quadrature. Figures 5 and 6 show reconstructed density matrix and Wigner function. Figure 2 and 5 suggest that both c.w. and pulsed lasers produce coherent states at given driving conditions above the lasing thresholds.

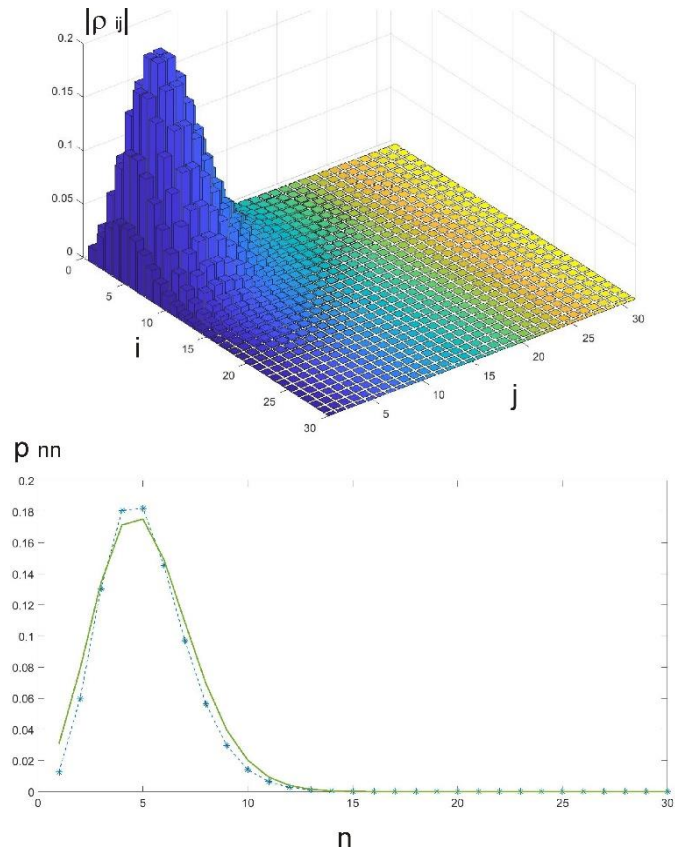

Figure 2. Absolute values of the density matrix and the photon number distribution of the 652 nm laser state. The blue line shows the experimental data, the green line represents a coherent state |α> with $|\alpha|^2 = 5$.

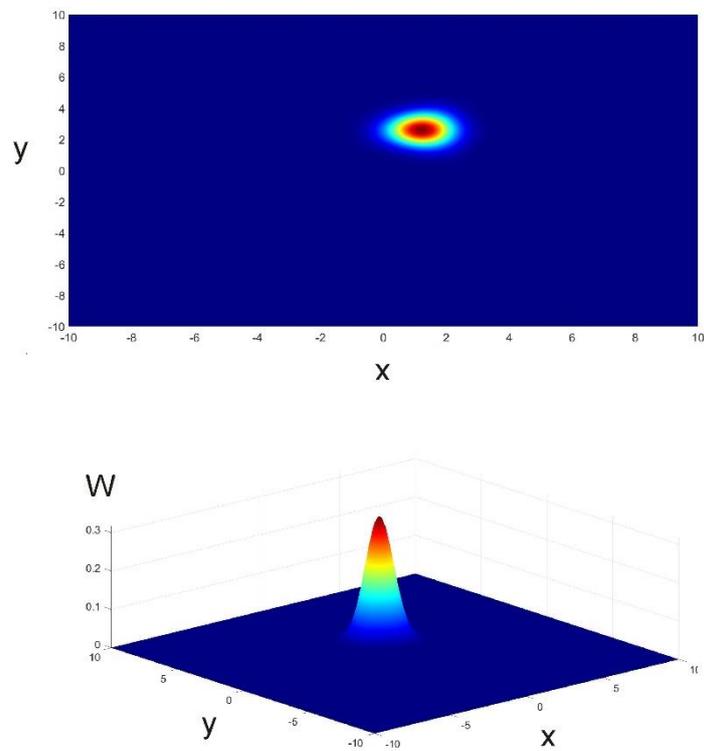

Figure 3. Reconstructed Wigner function of the 652 nm c.w. laser state.

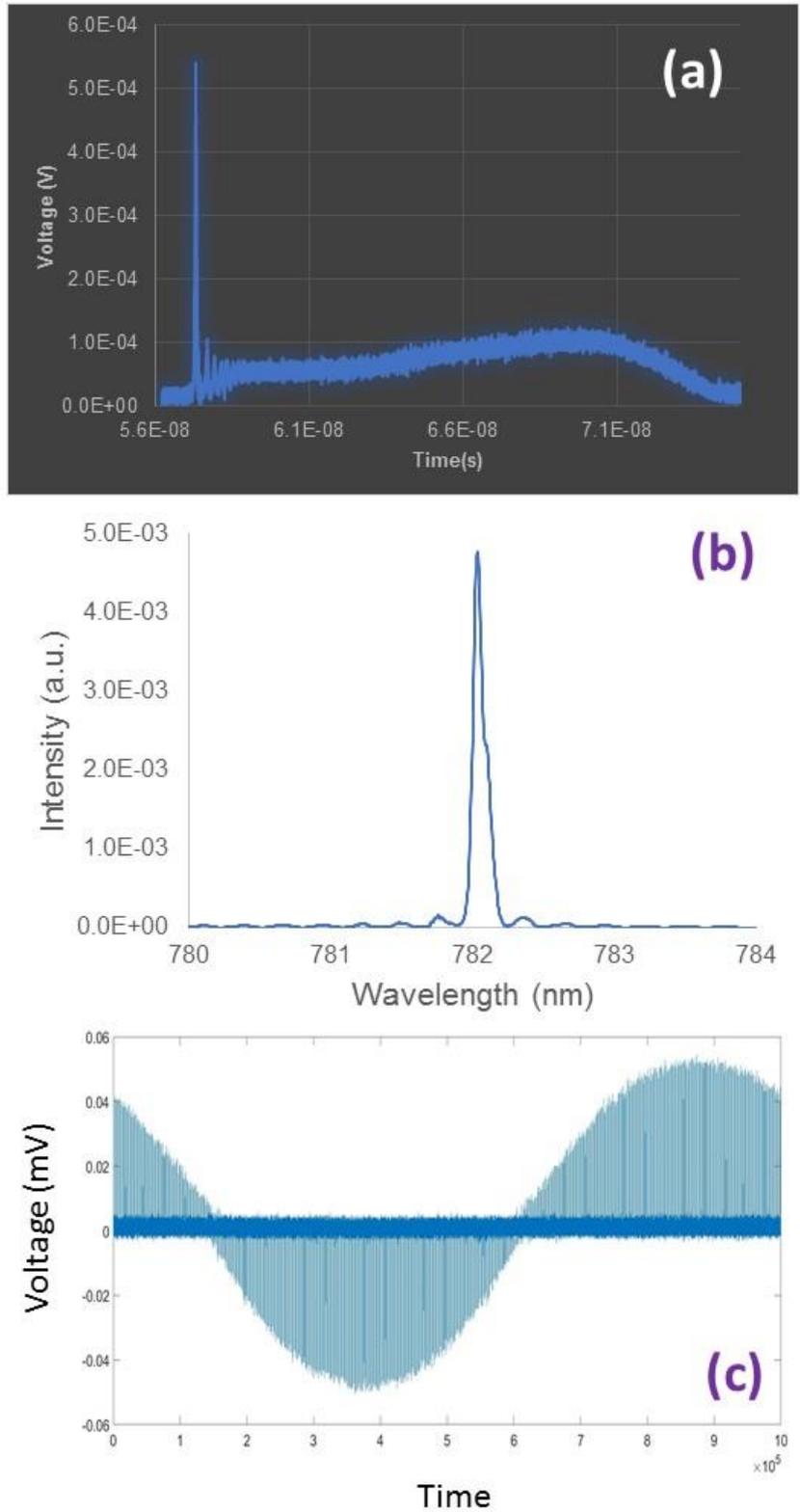

Figure 4. Output nanosecond pulse (a), the optical spectrum (b) and homodyne quadrature (c) of the second test laser. The pulse repetition rate is 2 MHz, the frequency of the PZT scanning is 2.5 kHz.

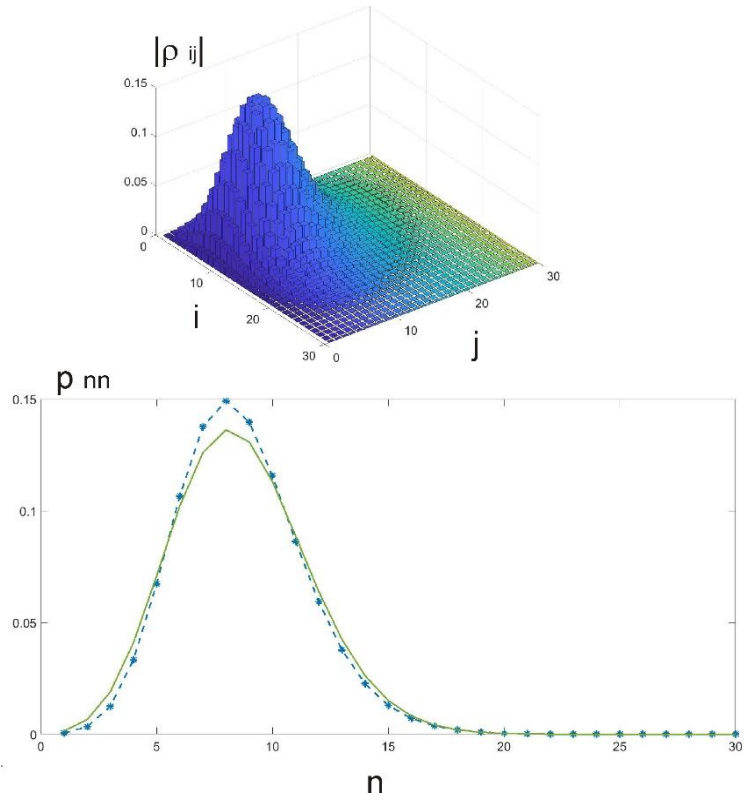

Figure 5. Absolute values of the density matrix and the photon number distribution of the 782 nm laser state. The blue line shows the experimental data, the green line represents a coherent state |α> with $|\alpha|^2 = 8$.

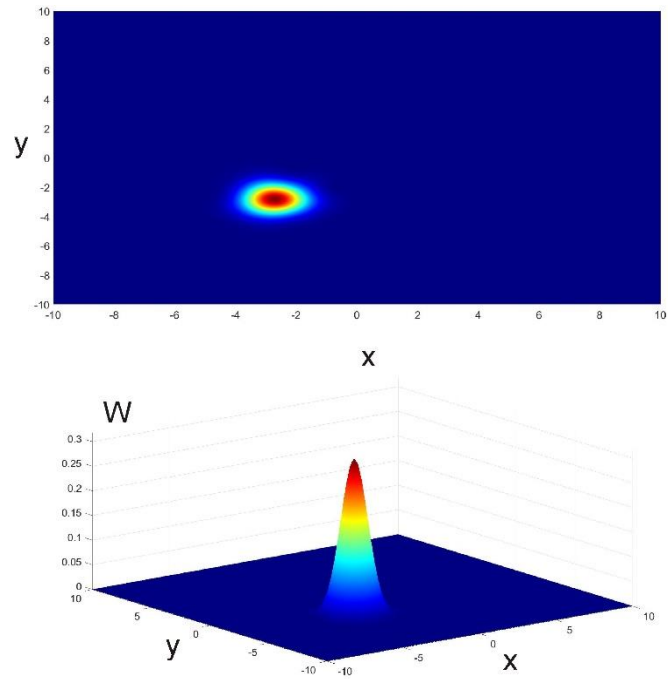

Figure 6. Reconstructed Wigner function of the 782 nm pulsed laser state.